\documentstyle[twoside,fleqn,espcrc2,epsf]{article}

\title{Review of Unquenched Results}
 
\author{Robert D. Mawhinney\address{Department of Physics, Columbia
  University, New York, NY 10027, USA}\thanks{Supported in part by
  the US Department of Energy and the RIKEN-BNL Research Center.}
}
 
\begin{document}

\def\thepage{CU--TP--965}
\thispagestyle{myheadings}

\begin{abstract}

One of the major frontiers of lattice field theory is the inclusion of
light fermions in simulations, particularly in pursuit of accurate,
first principles predictions from lattice QCD.  With dedicated
Teraflops-scale computers currently simulating QCD, another step
towards precision full QCD simulations is underway.  In addition to
ongoing staggered and Wilson fermion simulations, first results from
full QCD with domain wall fermions are available.  After some
discussion of work toward better algorithms, simulations completed to
date will be discussed.

\end{abstract}

\maketitle

\section{INTRODUCTION}

A major challenge in realizing the full potential of the
non-perturbative regularization provided by lattice field theory is the
presence of fermion fields in simulations.  In many cases, such as QCD
at finite density and the Hubbard model, the fermions require an
improvement in algorithms before large systems can be simulated.  For
QCD at zero chemical potential, simulations with dynamical fermions
will continue to become better controlled by faster computers, even
without (hoped for) theoretical improvements.

There is over a decade of experience with full QCD simulations,
although only recently has enough data become available for some
extrapolations to the continuum \cite{kenway_review}.  As with quenched
simulations questions of what are satisfactory volumes and lattice
spacings, what are reliable simulation lengths and what are the best
extrapolations need to be answered.  Removing the uncontrolled
truncation of the quenched approximation is vital to achieving
precision results, but answers to the above are very important.

After a brief overview of the work on algorithms presented at this
conference, I will focus on results for QCD, particularly the low lying
hadrons.  There is evidence the quenched hadron spectrum, in the
continuum and chiral limits, differs from nature
\cite{burkhalter_review}.  For full QCD, the light hadron spectrum is
interesting in its own right, provides the basis for a determination of
light quark masses and is a useful testing ground for exploring finite
volume effects and simulation lengths.  Also, hadronic states
differing only by their parity are degenerate unless chiral symmetry
is broken, so studying these other parity states in the spectrum
may show sensitivity to the inclusion of the quark determinant.

Most of the techniques developed in the quenched approximation for
measuring other observables can be easily adapted to full QCD
simulations.  Many groups are reporting on effects of dynamical
fermions outside of the light hadron spectrum ($f_B$, etc.).  Please
see the reviews of these areas for information.

\section{ALGORITHMS}

Three groups reported work on the algorithms used to produce the Markov
chain in numerical simulations.

\subsection{Meron Cluster Algorithm}

Wiese and collaborators (MIT) \cite{wiese}, have tackled the long
standing problem of algorithms for fermion systems where the weights in
the path integral are complex.  Computational times exponential in the
volume are needed if the weight is made real by taking its absolute
value and moving the phase to an observable.  By working with a
continuum Euclidean time formulation, where fermions are represented by
their world lines, their algorithms form clusters whose flip gives a
definite sign.  A second improvement insures that clusters which cancel
through a sign flip and those that don't are generated with similar
probabilities.  This algorithm, with its attendant improved estimator,
has allowed them to simulate various fermionic systems that are
intractable with conventional approaches.  Application of these ideas
to coupled gauge-fermion systems is being investigated.

\pagenumbering{arabic}
\addtocounter{page}{1}

\subsection{Truncated Determinant Approach}

Duncan, et.\ al.\ \cite{duncan} have been studying an algorithm where
the fermion determinant is split, in a gauge-invariant way, into an
infrared and ultraviolet part.  The ultraviolet part is modeled by an
effective action made up of small Wilson loops, while the infrared part
is determined precisely using a Lanczos procedure.  To date, they have
generated a sequence of gauge fields using just the infrared part of
the determinant and have found that the exact contribution of the
ultraviolet part is well fit by an expansion in a small number of
Wilson loops.  A somewhat surprising result also comes from the
generation of the gauge fields:  they first update links and then put
in a global accept/reject step based on the infrared part of the
determinant.  They find reasonable acceptance for this, even for large
volumes.

\subsection{Multiboson Algorithm}

de Forcrand and collaborators \cite{deForcrand} have been testing an
evolved version of Luscher's multiboson algorithm \cite{luscher} in a
real QCD context and comparing its performance to a standard, but
highly tuned, hybrid Monte-Carlo code of the SESAM collaboration.  This
new work uses only 24 boson fields for a simulation with $m_\pi/m_\rho
= 0.833$, a dramatic improvement.  This is achieved by an ``ultraviolet
filtering" of the determinant, where the determinant is filtered by an
exponential term made up of weighted small Wilson loops, and a
cancelling term enters the gauge action.  (This filtering is similar to
that used by \cite{duncan}, but here is part of an exact
algorithm.)  An optimized quasi-heatbath is then applied to the
combined boson-gauge system.  They conclude that for a $16^3 \times 24$
lattice, with $\beta = 5.6$ and the equivalent to $\kappa = 0.156$ for
Wilson fermions, the multiboson algorithm decorrelates the plaquette
better than HMC, measuring both algorithms in units of
applications of the Dirac operator.

\section{SIMULATION KEY}

To distinguish the different actions, the abbreviations below will be
used (following the CP-PACS collaboration).
\begin{center}
\begin{tabular}{cc}
gauge action & \\
P & plaquette \\
R & RG improved \\ \\
fermion action&  \\
S & staggered \\
W & Wilson \\
C & Clover \\
D & domain wall \\
\end{tabular}
\end{center}
The RG improved gauge action is the one proposed by Iwasaki, et.\ al.
\cite{iwasaki_rg}.

In the tables of simulations parameters, three results are generally
listed:  $m_\rho a$, $m_\pi/m_\rho$ and $m_\rho L$, where $L$ is the
spatial lattice extent and $a$ is the lattice spacing.  The value of
$m_\rho a$ (where available) is for the quark mass which gives the
quoted $m_\pi/m_\rho$.  Also note that a 770 MeV $\rho$ in a 3 fm box
has $m_\rho L = 11.6$.  Since these tables should only serve as a guide
to the general features of the data sets, errors are not listed and
only 2 significant figures are given.  Please see the original work for
more details.

The hybrid Monte Carlo (HMC) algorithm has been used for the 2 flavor
Wilson and domain wall and 4 flavor staggered simulations.  For 2
flavor staggered simulations the `R' algorithm of Gottlieb, et. al.\
\cite{gottlieb_r} has been employed.  The trajectory counts in the
tables are from the groups themselves and the definition of a
trajectory differs between the groups.

\section{STAGGERED FERMIONS}

Recent staggered simulations have been done by the MILC and Columbia
groups and are listed in Table \ref{tab:stagsim}.  MILC reported hadron
spectrum results last year \cite{gottlieb_lat98} and new results for
decay constants this year \cite{bernard}.  They found a continuum
extrapolation of $m_N/m_\rho$ for staggered fermions in full QCD gives
a larger value than in the quenched continuum.  The Columbia group has
new data for 2 and 4 flavor QCD with staggered fermions, with
simulations still underway \cite{sui}.

\begin{table}[htb]
\caption{Summary of staggered simulations.}
\label{tab:stagsim}
\begin{center}
\begin{tabular}{cccccc}
\multicolumn{6}{c}{ MILC - PS action - $24^3 \times 64$ - 2 flavors} \\
\hline\hline
      $\beta$
    & $m$
    & $m_\rho a$
    & $m_\rho L$
    & $m_{\pi}/m_{\rho}$
    & traj. \\ \hline
5.6	& 0.08 & 0.98 & 23  & 0.76 & 2000 \\
5.6	& 0.06 & 0.87 & 21  & 0.74 & 2000 \\
5.6	& 0.04 & 0.75 & 18  & 0.71 & 2000 \\
5.6	& 0.02 & 0.59 & 14  & 0.63 & 2000 \\
5.6	& 0.01 & 0.50 & 12  & 0.53 & 2000 \\
\\
\multicolumn{6}{c}
  {CU QCDSP - PS action - $16^3 \times 32$ - 2 flavors} \\
\hline\hline
      $\beta$
    & $m$
    & $m_\rho a$
    & $m_\rho L$
    & $m_{\pi}/m_{\rho}$
    & traj. \\ \hline
5.7	& 0.015 & 0.48 &  7.7	& 0.63 &  5000 \\
\\
\multicolumn{6}{c}
  {CU QCDSP - PS action - $16^3 \times 32$ - 4 flavors} \\
\hline\hline
      $\beta$
    & $m$
    & $m_\rho a$
    & $m_\rho L$
    & $m_{\pi}/m_{\rho}$
    & traj. \\ \hline
5.4	& 0.02 	& 0.50  & 8.0	& 0.71	& 5180 \\
5.4	& 0.015 & 0.48  & 7.7	& 0.67	& 4750\\
\\
\multicolumn{6}{c}
  {CU QCDSP - PS action - $24^3 \times 32$ - 4 flavors} \\
\hline\hline
      $\beta$
    & $m$
    & $m_\rho a$
    & $m_\rho L$
    & $m_{\pi}/m_{\rho}$
    & traj. \\ \hline
5.4	& 0.02 	& 0.49  & 12	& 0.72	& 5000 \\
5.4	& 0.01 	& 0.37  & 8.9	& 0.66	& 5000\\
\end{tabular}
\end{center}
\end{table}

\begin{figure}[htb]
\epsfxsize=\hsize
\epsfbox{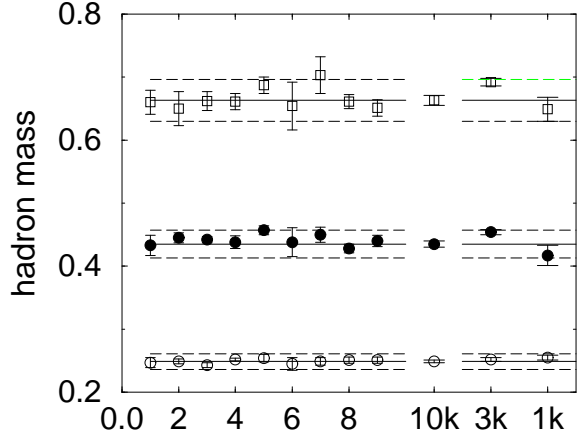}
\caption{ A comparison of hadron masses for a 10,000 trajectory
run (10k), a 3,000 trajectory run (3k) and a 1,000 trajectory
run (1k).  }
\label{fig:nf2_stat}
\end{figure}

There are existing results from 2 flavor staggered simulations at
$\beta = 5.7$ and $m = 0.01$ which give some information about run
lengths.  Figure \ref{fig:nf2_stat} shows, starting at the top, $m_N$,
$m_\rho$ and $m_\pi$ from the Columbia group \cite{dch} and Fukugita,
et.\ al.\ \cite{fukugita}.  The masses labeled ``1k" are from 1,000
trajectories on a $20^4$ lattice,  ``3k" from 3,000 trajectories on
$16^3 \times 32$ and  ``10k" from 10,000 trajectories on $16^3 \times
40$.  The points to the left of ``10k" are the 10,000 trajectory run
broken up in to nine 1,000 trajectory runs (plus thermalization)
\cite{dch_thesis}.  The solid horizontal lines are the ``10k" masses
and the dashed horizontal lines give $\pm5$\% of the central value.

The similar size for the errors on the 3k and 10k runs (the same
analysis was used for both) indicates that long correlation times were
likely missed in the 3,000 trajectory run.  The larger error bars for
the 1,000 trajectory runs may be due to short term noise effects.  For
these masses and couplings, achieving reliable errors at the few
percent level likely requires more than 10,000 trajectory run.

Another important question for full QCD simulations is the volume
required.  We now investigate this question through the splitting
between parity partners in the hadron spectrum.  We will see that, at
least for staggered fermions, this is a sensitive indicator of finite
volume effects.

\begin{figure}[htb]
\epsfxsize=\hsize
\epsfbox{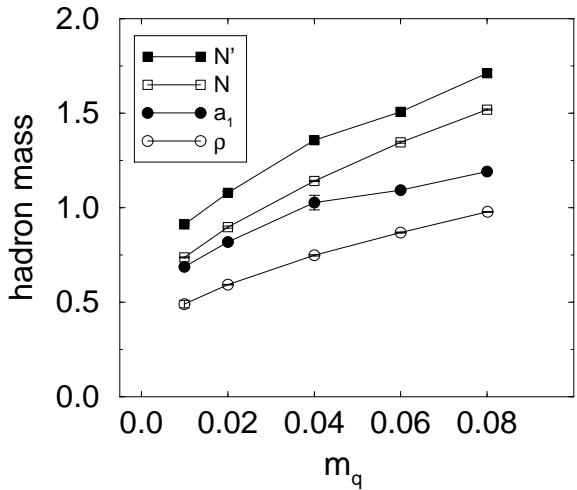}
\caption{ Some of the MILC $24^3 \times 48$, 2 flavor staggered
spectrum plotted versus quark mass.  The lines merely connect the
points.}
\label{fig:milc}
\end{figure}

MILC studies of finite volume effects in 2 flavor dynamical staggered
simulations for couplings up to $\beta = 5.6$ \cite{sg_tsu_97} show
lattice volumes of 2.5-3 fermi remove finite volume effects for
$m_\pi/m_\rho \sim 0.5$.  Their most recent $\beta=5.6$ results for a
$24^4 \times 48$ lattice \cite{sg_private} are shown in Figure
\ref{fig:milc}.  There is no sign of parity doubling, consistent with
their statement about finite volume effects.

\begin{figure}[htb]
\epsfxsize=\hsize
\epsfbox{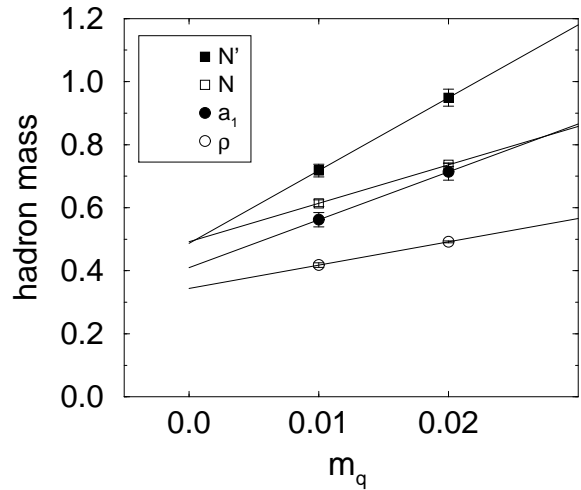}
\caption{Staggered 2 flavor simulations on $20^4$ lattices showing
parity doubling in $m_N$ and $m_{N^\prime}$ as $m \rightarrow 0$.}
\label{fig:fukugita}
\end{figure}

Fukugita, et.\ al. also studied finite volume effects but at weaker
coupling ($\beta = 5.7$).  Their data is plotted in Figure
\ref{fig:fukugita}, revealing parity doubling in the $m \rightarrow 0$
limit for the $N$ and $N^\prime$.   (The $N^\prime$ is the staggered
fermion parity partner of $N$.) Finite volume effects are likely
distorting the baryons, but not the mesons.  Which way $m_N$ and
$m_{N^\prime}$ move for larger volume is an open question, currently
being addressed by $24^3$ simulations underway at Columbia.

\begin{figure}[htb]
\epsfxsize=\hsize
\epsfbox{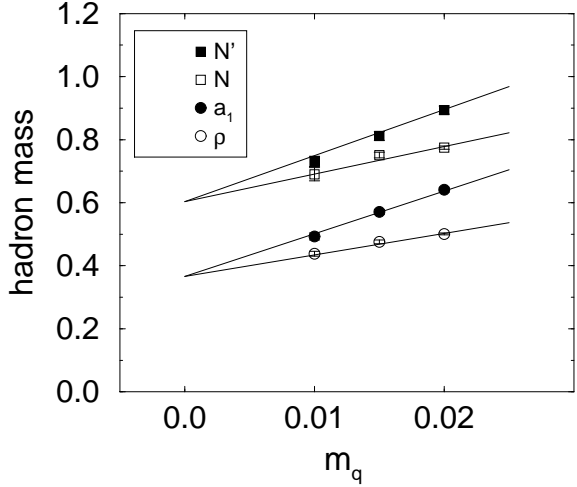}
\caption{Staggered 4 flavor simulations on $16^3 \times 32$ lattices
showing parity doubling in $m_N$, $m_{N^\prime}$ and $m_\rho$,
$m_{a_1}$ as $m \rightarrow 0$.}
\label{fig:cu_nf4_16}
\end{figure}

The Columbia group has previously reported parity doubling for 4 flavor
QCD on a $16^3 \times 32$ volume with $\beta = 5.4$.  (These parameters
give a rho mass within a few percent of the rho mass for 2 flavors on a
$16^3 \times 32$ lattice.)  Figure \ref{fig:cu_nf4_16} shows our
results, where the $m=0.01$ point is from the 256-node Columbia
machine, the $m=0.015$ point is from QCDSP and the $m=0.02$ result was
calculated on both the 256-node machine and QCDSP, which agree within
errors.  Parity doubling is clear for both mesons and baryons as $m
\rightarrow 0$.

\begin{figure}[htb]
\epsfxsize=\hsize
\epsfbox{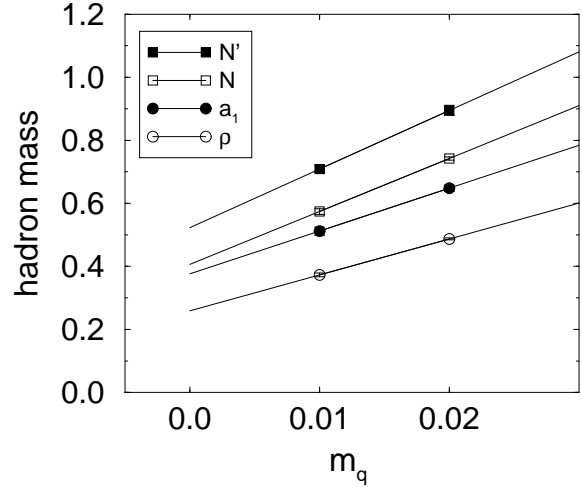}
\caption{Staggered 4 flavor simulations on $24^3 \times 32$ lattices
showing parity doubling eliminated for larger volumes.}
\label{fig:cu_nf4_24}
\end{figure}

The Columbia group has now completed a 5,000 trajectory simulation on a
$24^3 \times 32$ volume using QCDSP.  Figure \ref{fig:cu_nf4_24} shows
that the degeneracy between the hadrons in the $m \rightarrow 0$ has
gone away.  There is very little change in the $m=0.02$ masses, but for
$m=0.01$ $m_\rho$ and $m_N$ have dropped by $\sim 20$\% ($m_\rho$: $
0.438(8) \rightarrow 0.373(6)$, $m_N$: $ 0.690(21) \rightarrow
0.574(9)$).  This clearly shows finite volume effects distorting the
parity splittings and here it primarily effects the nucleon and rho
masses.  Also, given the large decrease in $m_\rho$, the larger $24^3$
lattice has $m_\rho L = 8.9$ very close to 7.0 for $16^3$.

Four flavors likely make this finite volume effect more pronounced, but
it is a warning for 3 flavor simulations.  We are currently simulating
with 2 flavors on a $24^3 \times 32$ volume to see how large the
effects are there.  Unfortunately, these large finite volume effects
are masking any information about the role of the determinant in the
parity splittings.

\begin{figure}[htb]
\epsfxsize=\hsize
\epsfbox{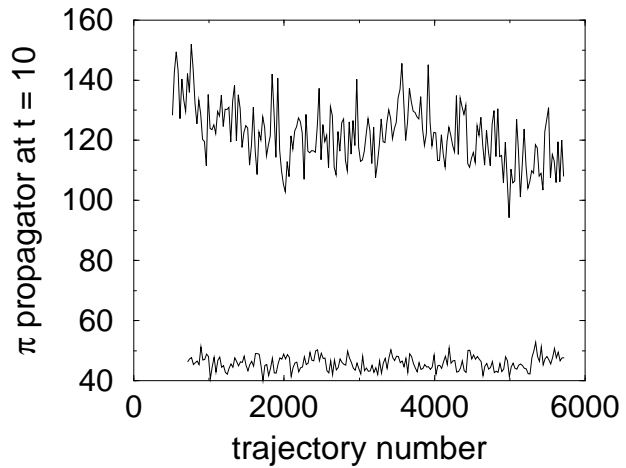}
\caption{The pion propagator at separation 10 for 4 flavor
simulations on $24^3 \times 32$ lattices plotted versus trajectory
number.  Much longer autocorrelation times seem visible for the upper
plot ($m=0.01$) when compared to the lower ($m=0.02$).}
\label{fig:pi_prop}
\end{figure}

A final message about run lengths from the 4 flavor staggered
simulations is shown in Figure \ref{fig:pi_prop}.  The upper line is
the pion propagator at a distance 10 lattice sites from the source for
the $m=0.01$ simulation, plotted against trajectory number.  The lower
line is the same propagator for the $m=0.02$ simulation.  The $m=0.01$
simulation shows fluctuations on a few thousand trajectory time scale.
This is clear evidence that very long runs are needed as the quark mass
is made smaller.  (Long autocorrelation times for topological charge
for 4 flavor staggered simulations on a $16^3 \times 32$ volume with
$\beta=5.35$ and $m=0.01$ have also been seen \cite{alles}.)

\section{WILSON FERMIONS}

The SESAM \cite{sesam}, UKQCD \cite{kenway_review} and CP-PACS
\cite{burkhalter_review} collaborations reported on full QCD
simulations last year and UKQCD \cite{garden} \cite{michael} and
CP-PACS \cite{kaneko} \cite{burkhalter} have new results this year.
The run parameters are given in Table \ref{tab:wilsim} and
\ref{tab:cppacs}.  Both UKQCD and CP-PACS are using clover improved
Wilson fermions;  UKQCD uses $C_{\rm SW}$ determined by the Alpha
collaboration and CP-PACS uses a tadpole improved value.

UKQCD chooses $\beta$ and $\kappa$ to keep the lattice spacing, as
determined using the Sommer scale $r_0$ and the static
quark--anti-quark potential, constant.  The physical volume then
corresponds to 1.7 fm for all lattice spacings they consider.  As
discussed above for staggered fermions, this volume can be expected to
be rather small.  They do report evidence that the potential at small
$r$ for the dynamical simulations lies below the value for quenched
simulations.  In addition, plotting vector meson masses against
pseudoscalar meson masses, they see a trend toward the physical
$(K,K^{*}$) value.

\clearpage


\begin{table}[htb]
\caption{Summary of UKQCD and SESAM 2 flavor Wilson fermion
simulations.}
\label{tab:wilsim}
\begin{center}
\begin{tabular}{cccccc}
\multicolumn{6}{c}
  {SESAM - PW action - $\beta = 5.6$ - $16^3 \times 32$} \\
\hline\hline
      \multicolumn{2}{l}{$\kappa$ }
    & $m_\rho a$
    & $m_\rho L$
    & $m_{\pi}/m_{\rho}$
    & traj. \\ \hline
\multicolumn{2}{l}{0.156}	& 0.53	& 8.5	& 0.83	& 5000 \\
\multicolumn{2}{l}{0.1565}	& 0.50	& 8.0 	& 0.81	& 5000 \\
\multicolumn{2}{l}{0.157}	& 0.46	& 7.4	& 0.76	& 5000 \\
\multicolumn{2}{l}{0.1575} 	& 0.41	& 6.6	& 0.68	& 5000 \\
\\
\multicolumn{6}{c} {UKQCD - PC action - $16^3 \times 32$} \\
\multicolumn{6}{c} {$\beta = 5.29$, 5.26, 5.2, 5.2, respectively} \\
\hline\hline
      $\kappa$
    & $c_{\rm sw}$
    & $m_\rho a$
    & $m_\rho L$
    & $m_{\pi}/m_{\rho}$
    & conf. \\ \hline
0.1340	& 1.92 & 0.70	& 11	& 0.83	& 101 \\
0.1345	& 1.95 & 0.65	& 10	& 0.78	& 101 \\
0.1350	& 2.02 & 0.59	& 9.4	& 0.69	& 150 \\
0.1355	& 2.02 & 	&	& 0.58	& 102 \\
\end{tabular}
\end{center}
\end{table}

\begin{table}[htb]
\caption{Summary of CP-PACS 2 flavor Wilson fermion simulations.}
\label{tab:cppacs}
\begin{center}
\begin{tabular}{cccc}
\multicolumn{4}{c}
  {CP-PACS - RC action - $\beta = 1.80$ - $12^3 \times 24$} \\
\hline\hline
      $\kappa$
    & $c_{\rm sw}$
    & $m_{\pi}/m_{\rho}$
    & traj. \\ \hline
0.1409 	& 1.60	& 0.81	& 6250 \\
0.1430 	& 1.60	& 0.75  & 5000 \\
0.1445 	& 1.60	& 0.70  & 7000 \\
0.1464 	& 1.60	& 0.55  & 5250 \\
\\
\multicolumn{4}{c}
  {CP-PACS - RC action - $\beta = 1.95$ - $16^3 \times 32$} \\
\hline\hline
     $\kappa$
    & $c_{\rm sw}$
    & $m_{\pi}/m_{\rho}$
    & traj. \\ \hline
0.1375 	& 1.53	& 0.80  & 7000 \\
0.1390 	& 1.53	& 0.75  & 7000 \\
0.1400 	& 1.53	& 0.69  & 7000 \\
0.1410 	& 1.53	& 0.59  & 7000 \\
\\
\multicolumn{4}{c}
  {CP-PACS - RC action - $\beta = 2.10$ - $24^3 \times 48$} \\
\hline\hline
      $\kappa$
    & $c_{\rm sw}$
    & $m_{\pi}/m_{\rho}$
    & traj. \\ \hline
0.1357 	& 1.47	& 0.81  & 2000 \\
0.1367 	& 1.47	& 0.76  & 2000 \\
0.1374 	& 1.47	& 0.69  & 2000 \\
0.1382 	& 1.47	& 0.58  & 2000 \\
\\
\multicolumn{4}{c}
  {CP-PACS - RC action - $\beta = 2.20$ - $24^3 \times 48$} \\
\hline\hline
      $\kappa$
    & $c_{\rm sw}$
    & $m_{\pi}/m_{\rho}$
    & traj. \\ \hline
0.1351 	& 1.44	& 0.80  & 2000 \\
0.1358 	& 1.44	& 0.75  & 2000 \\
0.1363 	& 1.44	& 0.70  & 2000 \\
0.1368 	& 1.44	& 0.63  & 2000 \\
\end{tabular}
\end{center}
\end{table}

The CP-PACS collaboration has been involved in extensive simulations of
full QCD with a variety of lattice spacings, quark masses and volumes.
Their parameter choices keep the spatial size fixed at $\sim 2.4$ fm,
using the $\rho$ mass at the physical $m_\pi/m_\rho$ value to set the
scale.  With their full QCD data set, they can extrapolate to the
continuum, with fixed finite volume effects.  To date, they only have
2,000 trajectories for their $\beta=2.10$ point, which experience with
staggered fermions suggests may not be long enough.  They are
addressing this issue.

\begin{figure}[htb]
\epsfxsize=\hsize
\epsfbox[50 300 550 720]{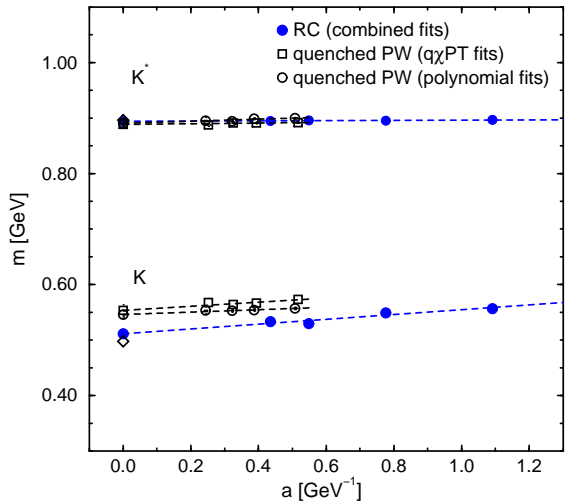}
\caption{ CP-PACS meson masses using the $\phi$ mass to set
the strange quark mass.}
\label{fig:meson.phi}
\end{figure}

CP-PACS found that a major difficulty with the quenched hadron spectrum
is the failure of a unique strange quark mass to give the physical
values for the $K$ and $\phi$ masses.  They have addressed this
question with their new data and one of their results is shown in
Figure \ref{fig:meson.phi}.  They find evidence that the $a \rightarrow
0$ extrapolation of the lattice result is much closer to the physical
$K$ mass ($\diamond$) than the quenched $a \rightarrow 0$ value.
Extrapolations of the octet and decuplet baryon masses \cite{kaneko}
have larger errors and definitive conclusions cannot be drawn.

\begin{figure}[htb]
\epsfxsize=\hsize
\epsfbox[50 100 550 550]{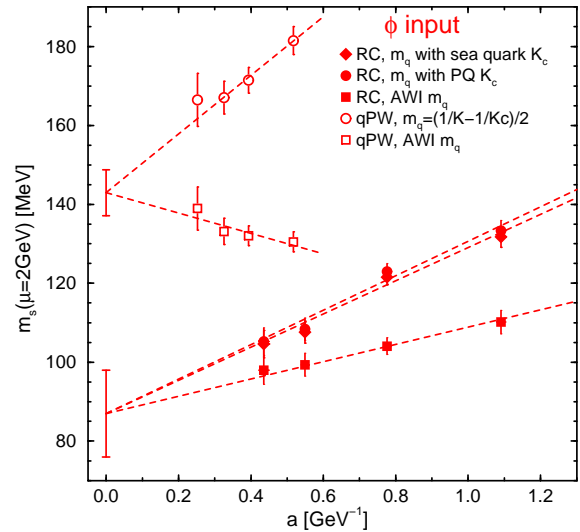}
\caption{ CP-PACS strange quark mass at $\mu = 2$GeV as determined
from the $\phi$ mass.}
\label{fig:ms.phi}
\end{figure}

Given their evidence that a single value of the strange quark mass
determines both the $K$ and $\phi$ masses, they have also determined
the $a \rightarrow 0$ strange quark mass at $\mu = 2$GeV.  One of their
results is shown in Figure \ref{fig:ms.phi}.  They find various ways of
determining the strange quark mass all agree in the $a \rightarrow 0$
limit and there is a large difference between the quenched and
unquenched values for this mass.  Their final result is $m_s = 87(11)$
MeV ($\phi$ input) and $m_s = 84(7)$ MeV ($K$ input).  These values are
considerably lower than those from phenomenology.  They also find
$m_{u,d} = 3.3(4)$ MeV.

The CP-PACS collaboration also reported a value for the flavor singlet
pseudo-scalar meson for 2 flavor QCD, referred to as the $\eta$
\cite{burkhalter}.  Using a volume source without gauge fixing (the
Kuramashi method) to measure the disconnected quark diagrams, the mass
difference between the $\pi$ and $\eta$ can be calculated, at least
when the disconnected diagrams are not far separated in time.  They
find $m_\eta$ non-zero in the chiral limit and a subsequent
continuum extrapolation gives $m_\eta = 863(86)$ MeV.

UKQCD has also reported a preliminary value for the $\eta$ mass
\cite{michael} for their set of dynamical lattices.  They
employ a variance reduction technique in finding the propagator for the
disconnected diagrams and get an $\eta$ mass around 800 MeV in the
chiral limit, with an uncontrolled systematic error.

Both groups see little problem with autocorrelation times for
these topologically sensitive measurements.  UKQCD finds an
autocorrelation time smaller than that found by the SESAM
collaboration.

\section{DOMAIN WALL FERMIONS}

The development of domain wall and overlap \cite{kaplan} \cite{shamir}
\cite{neuberger} formulations of lattice fermions has produced
a great deal of excitement for both analytical and numerical studies.
These formulations offer a way of separating the chiral limit from the
continuum limit and may lead to a lattice formulation of non-abelian
chiral gauge theories \cite{luscher_cgt}.  For QCD-like theories,
quenched simulations with domain wall fermions were first
done by \cite{blum-soni} and were reviewed last year
\cite{blum_review}.  Concurrently, simulations of the Schwinger model
\cite{vranas}, including dynamical fermions, showed that domain wall
fermions produced the expected physics and were compatible with
standard HMC algorithms.

Two areas where using fermions with better chiral properties are of
particular interest are in simulations studying QCD thermodynamics
\cite{vranas_thermo} and matrix elements
\cite{blum-soni-lat99} \cite{dawson} \cite{wingate}.  The character of
the QCD phase transition is governed by the chiral symmetries of the
theory.  For matrix element calculations, chiral symmetry can be vital
for controlling operator mixing and using chiral perturbation theory as
a guide.  Both of these areas are under active study with the QCDSP
computers, using domain wall fermions.

The Columbia group has been studying full QCD thermodynamics with
domain wall fermions.  The studies to locate the transition region and
set the parameters to use are detailed in \cite{vranas_thermo}.  As
part of this work, zero temperature scale setting calculations are also
required, which we will focus on here \cite{lingling}.  The full QCD,
zero temperature domain wall simulations done to date are detailed in
Table \ref{tab:dwfsim}.  For these dynamical simulations, a heavy
bosonic field, frequently called the Pauli-Villars field, is needed to
remove the bulk infinity that occurs when the extent of the fifth
dimension is sent to infinity.

\begin{table}[htb]
\caption{A summary of 2 flavor domain wall fermion simulations.
All simulations use a domain wall height of 1.9.}
\label{tab:dwfsim}
\begin{center}
\begin{tabular}{cccccc}
\\
\multicolumn{6}{c}
  {CU QCDSP - PD action - $\beta = 5.325$ - $8^3 \times 32$} \\
\hline\hline
      $m$
    & $L_s$
    & $m_\rho a$
    & $m_\rho L$
    & $m_{\pi}/m_{\rho}$
    & traj. \\ \hline
0.06	& 24  	&  1.3	& 10.4	& 0.64 & 1170 \\
0.02	& 24  	&  1.2	& 9.6	& 0.55 & 955 \\
\\
\multicolumn{6}{c}
  {CU QCDSP - PD action - $\beta = 5.325$ - $16^3 \times 16$} \\
\hline\hline
    $m$
    & $L_s$
    & $m_\rho a$
    & $m_\rho L$
    & $m_{\pi}/m_{\rho}$
    & traj. \\ \hline
0.02 & 24  	& 1.2	& 9.6	& 0.57 & 560 \\
\\
\multicolumn{6}{c}
  {CU QCDSP - RD action - $\beta = 1.9$ - $8^3 \times 32$} \\
\hline\hline
      $m$
    & $L_s$
    & $m_\rho a$
    & $m_\rho L$
    & $m_{\pi}/m_{\rho}$
    & traj. \\ \hline
0.02 & 24  	& 1.2	& 9.6	&  0.52 & 875 \\

\\
\multicolumn{6}{c}
  {CU QCDSP - RD action - $\beta = 2.0$ - $8^3 \times 32$} \\
\hline\hline
      $m$
    & $L_s$
    & $m_\rho a$
    & $m_\rho L$
    & $m_{\pi}/m_{\rho}$
    & traj. \\ \hline
0.06  & 24  	& 1.1	&  8.8	& 0.66  & 960 \\
0.02  & 24  	& 0.95	&  7.6	& 0.50  & 1010 \\
0.02  & 48  	& 1.0	&  8.0	& 0.43 	& 760 \\

\end{tabular}
\end{center}
\end{table}

Scale setting calculations to support thermodynamics studies
are necessarily at fairly strong coupling.  A first positive result
is that even for these course lattices, the HMC algorithm exhibits
no problems thermalizing and evolving lattices.  A clear region
where the Wilson line and the chiral condensate undergo a rapid
crossover \cite{vranas_thermo} is seen and scale setting calculations
have been done there.  An important question is how the residual
quark mass, $m_{\rm res}$ (due to mixing of the light modes between the
two walls), depends on the extent of the fifth dimension.

\begin{figure}[htb]
\epsfxsize=\hsize
\epsfbox{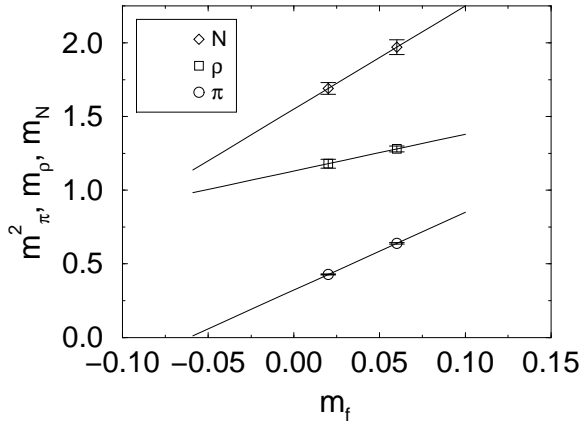}
\caption{Columbia QCDSP hadron masses for simulations with dynamical
domain wall fermions with the PD action.  The scale of the phase
transition on an $N_t = 4$ lattice determines the parameters.}
\label{fig:pd_hdm}
\end{figure}

For the PD action, the critical coupling on an $N_t = 4$ lattice
is $\beta = 5.325$, for $L_s = 24$, $m_f = 0.02$ and $M = 1.9$.
Figure \ref{fig:pd_hdm} shows hadron masses for this $\beta$
determined on $8^3 \times 32$ lattices for dynamical fermion
masses of 0.02 and 0.06.  By extrapolating $m_\pi^2$ to zero,
one finds $m_{\rm res} = 0.059(2)$.  (A similar value, within errors,
can also be extracted from the Ward-Takahashi identity \cite{gfleming}.)
This is clearly a large mass, compared to the input mass of 0.02.

Extrapolating the rho and nucleon mass to the point where $m_\pi^2$
vanishes gives $m_\rho = 1.02(7)$, $m_N = 1.14(9)$ and $m_\rho/m_N
= 1.10(12)$.  (The errors are calculated by simple propagation of
errors.)  While the length of the simulations ($\sim 1,000$
trajectories) may be sufficient at these strong couplings, only
a single volume and lattice spacing has been studied.  However,
it is encouraging that this ratio is much closer to its physical
value than for other fermion methods at this lattice spacing.

\begin{figure}[htb]
\epsfxsize=\hsize
\epsfbox{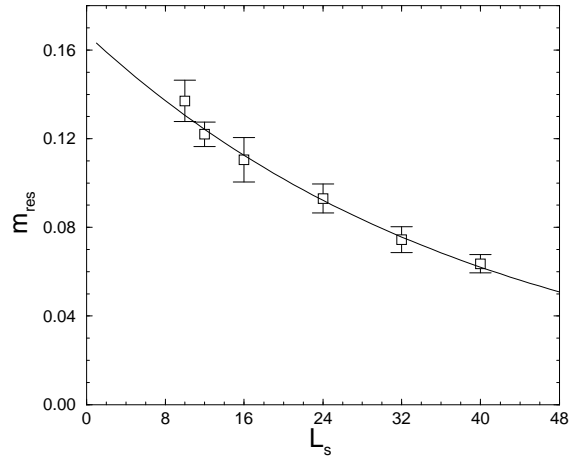}
\caption{$m_{\rm res}$ from the Ward-Takahashi identity as measured on
$8^3 \times 4 $ lattices in the confined phase with the PD action
and $\beta = 5.2$.}
\label{fig:pd_mres_ls}
\end{figure}

It is important to study the effects of increasing the fifth dimension
on the residual mass.  Figure \ref{fig:pd_mres_ls} shows the residual
mass versus $L_s$ (the length of the fifth dimension) for simulations
on $8^3 \times 4$ lattices with the PD action at $\beta = 5.2$ with a
quark mass of 0.02 and a domain wall height of 1.9.  These lattices are
in the confined phase.  The residual mass is determined from the axial
ward identity, as discussed in \cite{gfleming}.  The function shown is
$m_{\rm res}^{\rm (GMOR)} = 0.17(2) \times \exp ( -0.026(6) \times
L_s)$.  The data shows a vanishing residual mass in the large $L_s$
limit, but the falloff is very slow.  Increasing the fifth dimension by
24 drops the residual mass by about a factor of 2.

\begin{figure}[htb]
\epsfxsize=\hsize
\epsfbox{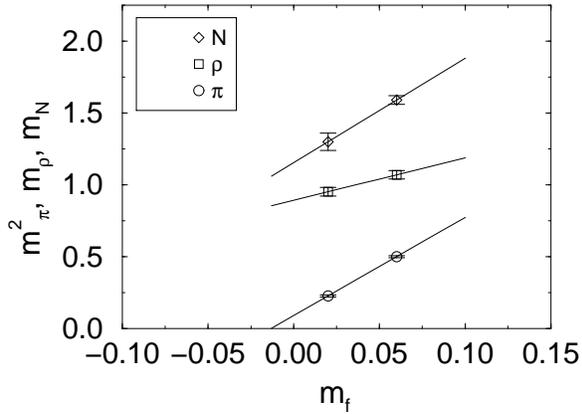}
\caption{Columbia QCDSP hadron masses for dynamical domain wall
simulations with the RD action for $\beta = 2.0$.}
\label{fig:rd_hdm}
\end{figure}

In an effort to decrease $m_{\rm res}$ for a fixed $L_s$, the Columbia
group has studied the renormalization group improved action proposed by
Iwasaki.  For quenched simulations, $m_{\rm res}$ is made smaller for a
given $L_s$, but the behavior with $L_s$ may not be improved
\cite{lingling}.  Results for hadron masses for dynamical simulations
with the RD action are shown in Figure \ref{fig:rd_hdm}.  These were
carried out on $8^3 \times 32$ lattices for $\beta=2.0$ with $L_s =
24$.  From the behavior of $m_\pi^2$ one finds $m_{\rm res} =
0.013(2)$.  At the point where $m_\pi^2$ vanishes $m_\rho = 0.855(49)$,
$m_N = 1.07(14)$ and $m_\rho/m_N = 1.25(14)$.  (Once again, these are
naive errors.)

While $m_{\rm res}$ is smaller for the RD simulations at $\beta = 2.0$
than the PD simulations at $\beta=5.325$, the physical lattice scales
are different.  For the RD action, the $N_t = 4$ phase transition is at
$\beta=1.9$.  We do not have two dynamical masses for the RD action at
$\beta=1.9$, but we can compare the pion masses at $m=0.02$.  For the
PD action at $\beta=5.325$, $m_\pi = 0.654(3)$ and for the RD action at
$\beta = 1.9$, $m_\pi = 0.604(3)$, a slight improvement.  (However, the
uncertainty in the determination of the critical temperature could
easily account for this difference.)

A direct comparison of the pion mass for the RD action ($\beta=2.0$,
$8^3 \times 32$) for $L_s =24$ gives $m_\pi = 0.475(7)$, while for $L_s
= 48$, $m_\pi = 0.420(10)$.  $m_{\rm res}$ is decreasing as $L_s$ is
increased, although the rate is slow.

\section{CONCLUSIONS}

With the current-Teraflops scale computers, full QCD simulations have
reached the point where the finite volume, long simulation time, small
quark mass and continuum limits can be probed, but not concurrently.
The HMC and HMD algorithms continue to be the techniques of choice, but
the multiboson algorithm has been evolved to a competitive level.
Staggered 4 flavor QCD shows large finite volume effects, even for
$m_\pi/m_\rho > 0.5$ and the parity splittings for light hadrons are
quite volume sensitive.  For currently accessible weaker coupling
simulations, runs of length greater than 10,000 trajectories are
probably needed.

CP-PACS has improved their results on the continuum limit of 2 flavor
QCD and find continuum meson masses closer to the physical values than
for the quenched case.  This leads them to a determination of the
strange quark mass which is lower than that expected
phenomenologically.  For staggered fermions, the new 2 flavor points
being produced by Columbia will augment the existing continuum
extrapolation of the MILC group.

Full QCD domain wall fermion simulations are straightforward,
but require a large fifth dimension at strong coupling.  It is
encouraging that on very coarse lattices $m_N/m_\rho$ is much lower
than for other fermion formulations.  Further studies will be needed to
check the scaling properties for domain wall fermions to see if
the use of coarser lattices can offset the extra calculational
cost of the fifth dimension.

\vspace{0.2in}
\noindent {\bf ACKNOWLDEGEMENTS}
\vspace{0.2in}

The author would like to thank R. Burkhalter, P. de Forcrand,
A. Duncan, J. Garden, S. Gottlieb, K. Kanaya, C. Michael, U. Wiese,
the CP-PACS collaboration and the MILC collaboration for 
providing access to their results.

\end{document}